\documentclass{aa}
\usepackage{hyperref}
\usepackage[varg]{txfonts} 
\usepackage{natbib}
\usepackage{ctable}
\usepackage{graphicx}
\usepackage{braket}
\begin{document}

\title{An IllustrisTNG View of the Caustic Technique for Galaxy Cluster Mass Estimation} 

\author{Michele Pizzardo \inst{\ref{1}}\thanks{\email{michele.pizzardo@smu.ca}}
\and Margaret J. Geller \inst{\ref{2}}
\and Scott J. Kenyon \inst{\ref{2}}
\and Ivana Damjanov \inst{\ref{1}}
\and Antonaldo Diaferio \inst{\ref{3},\ref{4}}
}

\institute{\label{1}Department of Astronomy and Physics, Saint Mary's University, 923 Robie Street, Halifax, NS-B3H3C3, Canada
\and \label{2}Smithsonian Astrophysical Observatory, 60 Garden Street, Cambridge, MA-02138, USA
\and \label{3}Dipartimento di Fisica, Universit\`a di Torino, via P. Giuria 1,  I-10125 Torino, Italy
\and \label{4}Istituto Nazionale di Fisica Nucleare (INFN), Sezione di Torino, via P. Giuria 1,  I-10125 Torino, Italy
} 

\date{Received date / Accepted date}

\abstract
{The TNG300-1 run of the IllustrisTNG simulations includes 1697 clusters of galaxies with $M_{200c}>10^{14}$M$_\odot$ covering the redshift range $0.01-1.04$. We build mock spectroscopic redshift catalogues of simulated galaxies within these clusters and apply the caustic technique to estimate the cumulative  cluster mass profiles. We compute the total true cumulative mass profile from the 3D  simulation data and calculate the ratio of caustic mass to total 3D mass, $\mathcal{F}_\beta$, as a function of cluster-centric distance and  identify the radial range  where $\mathcal{F}_\beta$ is roughly constant. \\
The filling factor, $\mathcal{F}_\beta=0.41\pm 0.08$, is constant on a plateau that covers a wide cluster-centric distance range, $(0.6-4.2)R_{200c}$. This calibration is insensitive to redshift. The calibrated caustic mass profiles are unbiased, with an average uncertainty of $23\%$. At $R_{200c}$, the average $M^C/M^{3D}=1.03\pm 0.22$; at $2R_{200c}$, the average $M^C/M^{3D}=1.02\pm 0.23$. Simulated galaxies are unbiased tracers of the mass distribution. IllustrisTNG is a  broad statistical platform for application of the caustic technique to large samples of clusters with spectroscopic redshifts for $\gtrsim 200$ members in each system. These observations will allow extensive comparisons with weak lensing masses and  will complement other techniques for measuring the growth rate of structure in the universe.  
}

\keywords{galaxies: clusters: general - galaxies: kinematics and dynamics - methods: numerical}

\maketitle

\section{Introduction}\label{sec:introduction} 

Massive clusters of galaxies serve as testbeds for cosmological models \citep[e.g.][]{Mantz08,Rapetti09,Vikhlinin09b,Mantz10a,Rapetti13,Mantz15} and as laboratories for galaxy evolution \citep{lin04,oliva14,Kravtsov18,sohn2019velocity,Golden22,Sohn2022bcg}. Clusters evolve over the age of the universe as they accrete material from their surroundings \citep[e.g.][]{press1974formation,white1978,bower1991,laceyCole93,sheth2002,zhang2008,corasaniti2011,desimone2011,achitouv2014,musso2018}. 

Measurements of the  mass profile of clusters beyond the virial radius  are critical for understanding the growth of clusters of galaxies \citep[e.g.,][]{diaferio2004outskirts,reiprich2013outskirts,Diemer2014,lau2015mass,walker2019physics,rost2021threehundred}. Radii typically larger than the fiducial radius $R_{200c}$\footnote{$R_{200c}$ is  the radius enclosing an average mass density 200 times  the critical density of the universe at the appropriate redshift.}, an estimate of the viral radius, probe the infall region of clusters where they still accrete mass. Mass profiles beyond the approximate virial radius enable estimates of the cluster accretion rate as a function of cosmological  epoch \citep{vandenbosch02,mcbride2009,Fakhouri2010,Diemer2017sparta2,deBoni2016,pizzardo2020,Pizzardo2022}. They also provide a route to estimating the splashback radius \citep[e.g.][]{adhikari2014,Diemer2014,More2015,Diemer2017sparta2,contigiani2019,Xhakaj2020}. 

At radii exceeding $\sim R_{200c}$ the cluster accretes galaxies and dark matter; the system is not in equilibrium \citep{ludlow2009,bakels2020}.  Techniques that assume dynamical equilibrium, including virial masses \citep{zwicky1937masses}, X-ray masses \citep{sarazin1988x}, and Jeans' analyses \citep{the1986jeans,merritt1987,binney2011galactic}, are not applicable at large radius.
Weak gravitational lensing \citep{bartelmann2010gravitational,hoekstra2013masses,Umetsu20essay} and the caustic technique \citep{Diaferio1997,Diaferio99,Serra2011} do not depend on dynamical equilibrium. Thus these techniques are a route to deriving cluster mass profiles at large radius. 

\citet{umetsu2014clash} and \citet{Umetsu16} obtain weak lensing mass profiles for radii $\lesssim (3-5.7)$~Mpc (a few times $R_{200c}$).
Steadily improving large datasets and sophisticated treatment of systematic issues \citep{Hoekstra2003,Hoekstra2011} continually advance the sensitivity and reliability of lensing-mass estimates that extend to larger radius \citep{Umetsu11,Umetsu13,Umetsu20essay}.

The caustic technique \citep{Diaferio1997,Diaferio99} is another strategy for mass estimation beyond $R_{200c}$. This dynamical method exploits the trumpet-like pattern in the projected phase-space distribution of cluster galaxies that results from the continual infall of matter. The caustic pattern reflects the escape velocity from the cluster. Identification of the region where the phase space density changes sharply enables reconstruction of the mass profile. 

Collisionless N-body simulations suggest that the caustic  mass profile is an unbiased estimator with a reliability of $\sim$ 50\% for radii $\lesssim 4R_{200c}$. Application of the caustic technique requires dense spectroscopic sampling of a cluster
including more than $\sim 150$ galaxies within a radius of $\sim 3R_{200c}$ \citep{Diaferio99,Serra2011}.  

\citet{Rines2006CIRS} and \citet{Rines2013HeCS} use two large spectroscopic surveys as a basis for application of the caustic technique to well-defined sets of massive clusters. The  Cluster Infall Regions in the Sloan Digital Sky Survey (CIRS) and the Hectospec Cluster Survey (HeCS) surveys characterize the mass profiles of $\sim 130$ clusters of galaxies within a limiting radius of $\sim 5$~Mpc. \citet{pizzardo2020} use the caustic technique to estimate the mass accretion rate of these clusters based on the  mass profile in the range $\sim (2-3) R_{200c}$. \citet{Pizzardo2022} extend the study to stacked clusters from the HectoMAP spectroscopic survey \citep{sohn2021hectomap,sohn2021cluster}. The accretion  rates agree with the predictions of N-body simulations of the $\Lambda$CDM model.

Weak lensing and the caustic technique offer complementary approaches for studying the outskirts of clusters of galaxies.  \citet{diaferio2005caustic} and \citet{Geller_2013} show that  caustic and weak lensing profiles of the $\sim 20$ HeCS clusters agree within $20-30\%$.
Future spectrographs \citep[see, e.g.,][]{Dalton12,Tamura16,MSE19} will enable larger spectroscopic surveys to compare with the ever increasing number of high quality weak lensing mass profiles derived from comprehensive, deep photometric surveys.

For application to future ambitious spectroscopic surveys combined with weak lensing, a broad, robust platform for the caustic technique is needed. Previous studies calibrate the technique with N-body simulations where the galaxies were either identified  with semi-analytic prescriptions  \citep{Diaferio99} or associated with random samples of dark matter particles \citep{Serra2011}. 
Semi-analytical prescriptions for galaxy formation do not capture the full complexity of the hydrodynamics of clusters and dark matter only simulations rely on the assumption of negligible velocity bias between galaxies and dark matter. In addition, previous work was  
limited to nearby clusters with $z \approx 0$.

We use a large sample of clusters from the magneto-hydrodynamical IllustrisTNG simulations \citep{Pillepich18,Springel18,Nelson19} to explore and to calibrate the caustic technique in a platform that provides the phase-space distributions of cluster galaxies and dark matter for the same systems. The analysis of simulated galaxy catalogs makes no assumptions about cluster dynamics. IllustrisTNG provides a new, broad statistical platform for application of the caustic technique for systems with $z \lesssim 1$.

In Sect. \ref{sec:ct}, we review  the caustic technique.  Sect. \ref{sec:simulated} describes the IllustrisTNG simulations, the sample of simulated clusters, and the construction of mock galaxy redshift surveys of the 1697 clusters in the IllustrisTNG sample. Sect. \ref{sec:fb} outlines the calibration of the caustic technique in the redshift range $0.01-1.04$. Sect. \ref{sec:performance} evaluates the calibrated caustic technique as an estimator of  cluster radius and mass. Sect. \ref{sec:discussion} compares the simulated galaxies with the dark matter as tracers of the total matter distribution. We also include a survey of some previous applications of the caustic technique. We conclude in Sect. \ref{sec:conclusion}

\section{The Caustic Technique}
\label{sec:ct}

The caustic technique \citep[][]{Diaferio1997,Diaferio99,Serra2011,Serra2013} estimates the three-dimensional cumulative mass profile (from now on, the ``caustic mass profile'') of a cluster from the line-of-sight escape velocity profile of the cluster members, $v_{\rm esc,los}(r)$, where $r$ is the projected cluster-centric distance.  

The $r-v_{\rm los}$ diagram, the line-of-sight velocity with respect to the cluster center, $v_{\rm los}$, as a function of $r$, is the basis for the caustic mass profile. In this space, cluster galaxies delineate a  trumpet-shaped pattern with a decreasing amplitude as $r$ increases. The gravitational potential of the cluster causes departure of infalling galaxies from the Hubble flow thus producing this  distinctive pattern. Throughout this region, the cluster is not in dynamical equilibrium.

We define the caustics as the symmetric boundaries of the trumpet-shaped region of the $r-v_{\rm los}$ diagram. The caustic amplitude, $\mathcal{A}(r)$, is half of the distance between the upper and lower caustic. The caustic technique locates the caustics and computes $\mathcal{A}(r)$.
\citet{Diaferio1997} showed that the caustic amplitude approximates the escape velocity profile from a cluster, $\mathcal{A}(r)\approx v_{\rm esc,los}(r)$. The square of the caustic amplitude, $\braket{v_{\rm esc,los}^2(r)}$, is linked to the gravitational potential of the cluster and thus to its mass profile.  
The caustics separate member galaxies that lie between the upper and the lower caustic from foreground/background objects.

The caustic technique estimates the mass profile as
\begin{equation}\label{eq:CT}
	GM (<r) = {\cal F}_\beta \int_{0}^{r} {\cal A}^2(R) \,{\rm d}R ~ ,
\end{equation}
where ${\cal F}_\beta$ is a constant filling factor.
In the original formulation of the caustic technique, ${\cal F}_\beta$ is the average of a function that combines the mass density profile $\rho(r)$, the gravitational potential $\phi(r)$, and the velocity anisotropy parameter $\beta(r)$. In hierarchical clustering scenarios, this function depends only weakly  on the cluster-centric distance for $r \gtrsim 0.5R_{200c}$ \citep{Diaferio99,Serra2011}. 

Previous studies calibrated ${\cal F}_\beta$ based on collisionless N-body simulations at $z=0$. 
This approach neglects any bias between dark matter particles and galaxies. Previous studies did not investigate the dependence of ${\cal F}_\beta$ on redshift.  Here we calibrate ${\cal F}_\beta$ with the IllustrisTNG simulations in the redshift range $(0.01-1.04)$.

\section{The IllustrisTNG Simulations}
\label{sec:simulated}

We extract simulated clusters from the TNG300-1 run of the IllustrisTNG simulations \citep{Pillepich18,Springel18,Nelson19}. This sample includes  three-dimensional (3D) matter distributions and galaxy mock redshift surveys of the clusters at eleven different redshifts. 
We describe the simulation in Sect. \ref{subsec:tng}, the samples of 3D clusters in Sect. \ref{subsec:3d},  and the galaxy mock redshift surveys in Sect. \ref{subsec:mockgal}. 

\subsection{Basic properties of the IllustrisTNG TNG300-1 Simulation} 
\label{subsec:tng}

We extract cluster samples from the IllustrisTNG simulations \citep{Pillepich18,Springel18,Nelson19}, a set of gravo-magnetohydrodynamical simulations based on the $\Lambda$CDM model. Each simulation differs in the size of the simulated volume, the resolution, and the matter content.  

The simulations are normalised at $z=127$ with the Planck cosmological parameters \citet{ade2016planck}: cosmological constant $\Omega_{\Lambda 0}=0.6911$, cosmological dark matter density $\Omega_{m0}=0.3089$, baryonic mass density $\Omega_{b0}=0.0486$, Hubble constant $H_0=67.74$~km~s$^{-1}$~Mpc$^{-1}$, power spectrum normalisation $\sigma_8=0.8159$, and power spectrum index $n_s=0.9667$. 
All of  the IllustrisTNG baryonic runs are based on the AREPO code \citep{Springel10} which solves the equations of continuum magnetohydrodynamics coupled with Newtonian self-gravity. The simulations include the following baryonic processes: primordial and metal-line cooling in the presence of an ionizing background radiation field, stochastic star formation, stellar evolution with the associated chemical enrichment and mass loss,  ISM pressurization resulting from unresolved supernovae, stellar feedback, seeding and growth of supermassive black holes, feedback from supermassive black holes, and the dynamical impact of the amplification of a small primordial magnetic field.   

We use the TNG300-1 run of IllustrisTNG. This baryonic run has the highest resolution among the runs with the largest simulated volumes. The  comoving box size is $302.6$~Mpc on a side. 
TNG300-1 contains $2500^3$ dark matter particles with mass $m_{\rm DM} = 5.88 \times 10^{7}~\rm{M_{\odot}}$ and the same number of gas cells with average mass $m_b = 1.10\times 10^{7}~\rm{M_{\odot}} $.

Structures in the simulations are identified with a Friends-of-Friends (FoF) algorithm with linking length $\lambda=0.2\bar{d}$, where $\bar{d}$ is the mean Lagrangian inter-particle separation. The algorithm is applied only to the dark matter particles. Gas, stars, and black holes are then attached to the same FoF group as their nearest dark matter particle. 

The substructures of each FoF group are identified as the gravitationally bound structures within the group by means of the Subfind algorithm \citep{subfind2001}, which  runs over all the particle types. A synthetic cluster corresponds to a cluster-mass FoF group. The cluster member galaxies correspond to the substructures identified within the group. 
We identify the  center of each cluster as the center of mass of its most massive, or primary, Subfind group. The center of mass is the sum of the mass-weighted coordinates of all the particles and cells in the substructure. 

\subsection{3D information}\label{subsec:3d}

We build the sample of synthetic clusters starting from the group catalogues compiled by the TNG Collaboration. These catalogues list the global properties of the FoF groups along with the substructures identified by the Subfind algorithm. We use these catalogues to select all of the FoF groups with $M_{200c,FoF}^{3D} > 10^{14}$M$_\odot$. 

For each  FoF halo, we extract a spherical volume from the raw snapshots. This volume is centered on the center of mass of the most massive substructure of the halo.  The radius  of the volume is $10R_{200c,FoF}^{3D}$, and it contains all of the matter species including the dark matter, gas, stars, and black holes. 
From the 3D distribution of matter, we compute the true cumulative mass profile (from now on, the ``true mass profile'') of each cluster, $M^{3D}(r)$, in 200 logarithmically spaced bins in the radial range $(0.1-10)R_{200c,FoF}^{3D}$. From these profiles, we compute the $R_{200c}^{3D}$ and $M_{200c}^{3D}$ for each cluster. We use $M_{200c}^{3D}$ to select the final  cluster samples of the systems with mass $M_{200c}^{3D} > 10^{14}$~M$_\odot$. We include clusters in eleven redshift intervals: $z=0.01, 0.11, 0.21, 0.31, 0.42, 0.52, 0.62, 0.73, 0.82, 0.92$, and $1.04$.

Table \ref{table:3dinfo} lists the number of synthetic clusters, the medians and the 68\% widths of the distributions of their masses ($M_{200c}^{3D}$), and the minimum and the maximum $M_{200c}^{3D}$ at each redshift.

\begin{table*}[htbp]
\begin{center}
\caption{\label{table:3dinfo} Cluster samples from Illustris TNG300-1}
\begin{tabular}{cccccccc}
\hline
\hline
 $z$ & no of clusters & median $M_{200c}$ & $68\%$ range & min $M_{200c}$ & max $M_{200c}$ & no of galaxies &$68\%$ range   \\
     & & $\left[10^{14}~\text{M}_{\odot }\right]$  & $\left[10^{14}~\text{M}_{\odot }\right]$ & $\left[10^{14}~\text{M}_{\odot }\right]$  & $\left[10^{14}~\text{M}_{\odot }\right]$ &  within $3R_{200c}^{3D}$ &\\
\hline
 & & \\
0.01 & 282 & 1.59 & 1.13-3.01 & 1.00 & 15.0 & 172 & (116-316) \\
0.11 & 255 & 1.58 & 1.13-2.97 & 1.00 & 12.6 & 173 & (122-316) \\
0.21 & 231 & 1.47 & 1.16-2.84 & 1.01 & 12.3 & 173 & (128-313) \\
0.31 & 201 & 1.50 & 1.14-2.66 & 1.01 & 13.6 & 182 & (130-309) \\
0.42 & 178 & 1.43 & 1.12-2.42 & 1.00 & 13.6 & 181 & (132-301) \\
0.52 & 145 & 1.41 & 1.14-2.47 & 1.01 & 12.1 & 190 & (134-300) \\
0.62 & 122 & 1.43 & 1.11-2.43 & 1.00 & 8.84 & 189 & (135-287) \\
0.73 & 98 & 1.35 & 1.10-2.37 & 1.00 & 8.92 & 191 & (146-335) \\
0.82 & 80 & 1.38 & 1.08-2.38 & 1.00 & 8.43 & 202 & (147-361) \\
0.92 & 60 & 1.38 & 1.11-2.42 & 1.00 & 7.60 & 210 & (158-357) \\
1.04 & 45 & 1.43 & 1.13-2.01 & 1.02 & 4.37 & 207 & (174-352) \\

\hline
 \end{tabular}
 \end{center}
 \end{table*}

\subsection{Mock redshift surveys}\label{subsec:mockgal}

We associate a galaxy mock redshift survey with each simulated cluster. We follow a procedure similar to  \citet{pizzardo2020}. They build mock catalogues of clusters from the simulated 3D distribution of the dark matter particles of an N-body simulation.
Here we use the distribution of the synthetic galaxies rather than the dark matter particles. This approach produces mock catalogues that automatically include any velocity or spatial bias between dark matter particles and galaxies.

To generate each mock catalogue, we extract a squared-basis truncated pyramid centred on the cluster. The smaller basis is  closer to the fictitious observer and the pyramid axis is aligned with the $x$-axis of the simulation that we identify with the line of sight \citep[see Fig. 1 of][]{pizzardo2020}. The height of the simulated pyramid is $2b_L\approx 177$~Mpc. The vertical section of the pyramid at the center is a square with $r_{FOV}\approx 17.7$~Mpc on a side. 

We use the group catalogues to select all of the substructures with a center of mass within the pyramid. 
To simulate catalogues of cluster galaxies, we consider only the Subfind substructures with stellar mass  $ > 10^8~$M$_\odot$. This selection mimics observable galaxies. Hereafter we refer to these substructures as galaxies.

We use these mock catalogues to estimate the caustic mass profile of the  clusters. The basis of this technique is the $r-v_{\rm los}$ diagram (Sect. \ref{sec:ct}). To transform the simulations to this observable plane, we need the positions on the sky and the redshifts of the galaxies. Thus, for each simulated cluster, we translate the 3D galaxy coordinates into right ascension $\alpha$, declination $\delta$, and total redshift $z$ along the line-of-sight.  

We translate the original three-dimensional comoving position, $\mathbf{r}_{c,i}$, of each galaxy in the pyramidal volume so that the comoving distance between the observer and the centre of the pyramid is $r_s = c/H_0 \int_{0}^{z_s} dz^\prime/E(z)$, where $E(z)=[(\Omega_{m0}+\Omega_{b0})(1+z)^3+\Omega_\Lambda]^{1/2}$ in the flat $\Lambda$CDM model, and $z_s$ is the redshift of the particular snapshot. The new 3D positions are $\mathbf{r}_i = \mathbf{r}_{s}+\mathbf{r}_{c,i}$. Setting the celestial coordinates of the center of the pyramid to $(\alpha_{c}, \delta_c)=(\pi/2, 0)$ in radians, standard geometrical transformations yield the celestial coordinates $(\alpha_i,\delta_i)$ of the synthetic galaxies. The observed redshift associated with each galaxy is $cz_{\rm obs} = cz_i + v_{\rm los}(1+z_i)$, where $z_i$ is the cosmological redshift obtained by inverting the integral expression of the comoving distance between the observer and the galaxy, $r_i= c/H_0 \int_{0}^{z_i} dz^\prime/E(z)$, and $v_{\rm los}$ is the component of the peculiar velocity of the galaxy along the observer's line of sight. Each mock catalogue lists the two celestial coordinates and the observed redshift of the constituent galaxies. 

The median number of galaxies in the catalogues is 1835, with a  68\% range $\sim(1350,2430)$. The catalogues include foreground and background objects. Within a three-dimensional distance of $3R_{200c}^{3D}$, the median number of galaxies in the catalogues lies in the range $172-210$ (Table \ref{table:3dinfo}, column 7). Previous studies  suggest that this sampling is a solid basis for application of the caustic technique \citep{Serra2011}.

\section{Calibration of the caustic technique}\label{sec:fb}

The IllustrisTNG simulations provide a platform for calibrating the caustic technique.
The  mock catalogues extracted from the TNG300-1 simulation allow the first measurement of $\mathcal{F}_\beta$ (Eq. \ref{eq:CT}) based on the distribution of simulated galaxies rather than dark matter particles. TNG300-1 also enables the first investigation of possible redshift dependence of $\mathcal{F}_\beta$ in the range (0.01-1.04).

Sect. \ref{subsec:caugal} describes  the  application of the caustic technique to the mock catalogues. Sect. \ref{subsec:fbeta} describes the measurement of  filling factor, $\mathcal{F}_\beta$, and its behavior as a function of redshift.

\subsection{Caustic mass profiles from the mock catalogues}\label{subsec:caugal}

We apply the caustic technique to  1697 mock catalogues to obtain a set of uncalibrated  cumulative  caustic mass profiles. Each mass profile is the radial integral of the square of the caustic amplitude (Eq. \ref{eq:CT}).
The caustic technique uses a hierarchical binary tree based on projected binding energy to select the galaxies used to build the $r-v_{\rm los}$ diagram. The technique is based on the cluster-centric distances and line-of-sight velocities of the galaxies relative to the angular coordinates and redshift of the cluster center. The center of each synthetic cluster is the position of the most massive substructure within its corresponding FoF halo, analogous to choosing the BCG as the cluster center \citep{Sohn2022bcg}. The position of this  substructure is consistent with the center of mass of the FoF halo.

To apply the caustic technique, we smooth the data in the $r-v_{\rm los}$ diagram to construct a continuous distribution. The parameter $h_c$ determines the smoothing scale used to build the continuous density distribution \citep[see, e.g., Eqs. 15-17 of][]{Diaferio99}. We  adopt no constraint on the value of the smoothing parameter $h_c$. The standard approach that we follow  sets $h_c$ with an adaptive kernel that minimizes the integrated square error between the continuous density estimator and the true density determined by the  $r-v_{\rm los}$ diagram \citep[see Eq. 18 of][]{Diaferio99}. Finally the caustic technique  locates the caustics as isocurves of the continuous 2D galaxy number density in phase space. 

\begin{figure}
    \centering
    \includegraphics[width=\columnwidth]{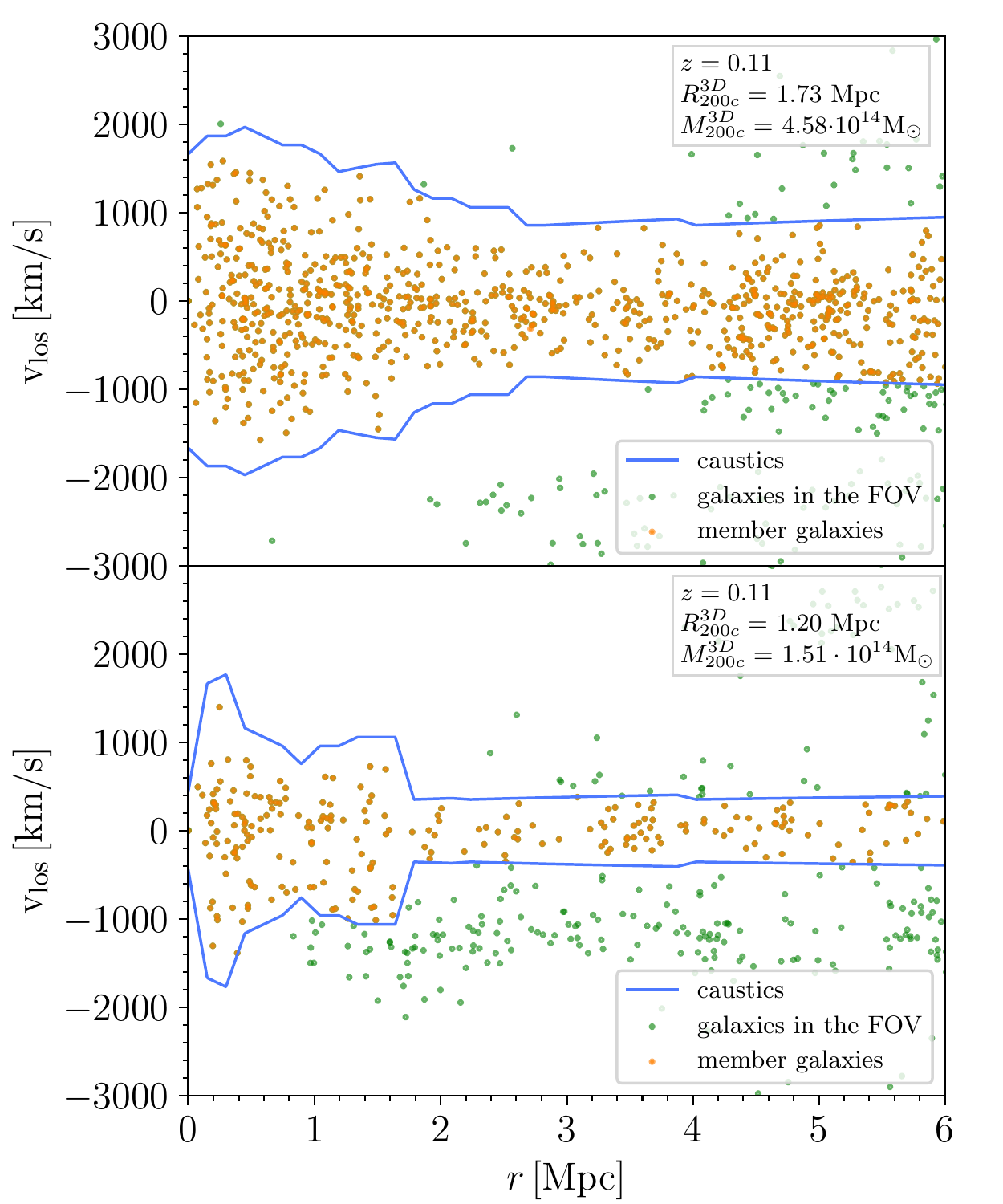}
    \caption{Two examples of simulated  $r-v_{\rm los}$ diagrams at $z=0.11$. Blue curves show the caustic profiles. Points denote simulated cluster members (orange) and background/foreground galaxies (green).}
    \label{fig:comp90gal}
\end{figure}

Figure \ref{fig:comp90gal} shows two examples of typical $r-v_{\rm los}$ diagrams at $z = 0.11$ with the associated caustic profiles. The upper (lower) panel shows the result for a cluster with mass larger (smaller) than the median mass. In each panel, points represent  simulated galaxies. The blue curves show the caustic profiles. The cluster members delineate a trumpet-shaped pattern in the $r-v_{\rm los}$ diagram (Sect. \ref{sec:ct}). The caustic technique locates the caustics which delimit this region. The caustic curves separate  cluster members  (orange points) from foreground/background objects (green points). The cluster in the upper (lower) panel has 710 (213) caustic member galaxies within $R_{200c}^{3D}$.

\subsection{Measurement of the filling factor}\label{subsec:fbeta}

Calibration of the caustic technique with the TNG300-1 simulation has two main steps. First, we identify the radial range where the ratio between the caustic and the true mass profiles is approximately constant. Then, the  typical ratio between the two masses on this plateau calibrates the filling factor that normalizes the mass ratio to unity.

\begin{figure}
    \centering
    \includegraphics[scale=0.73]{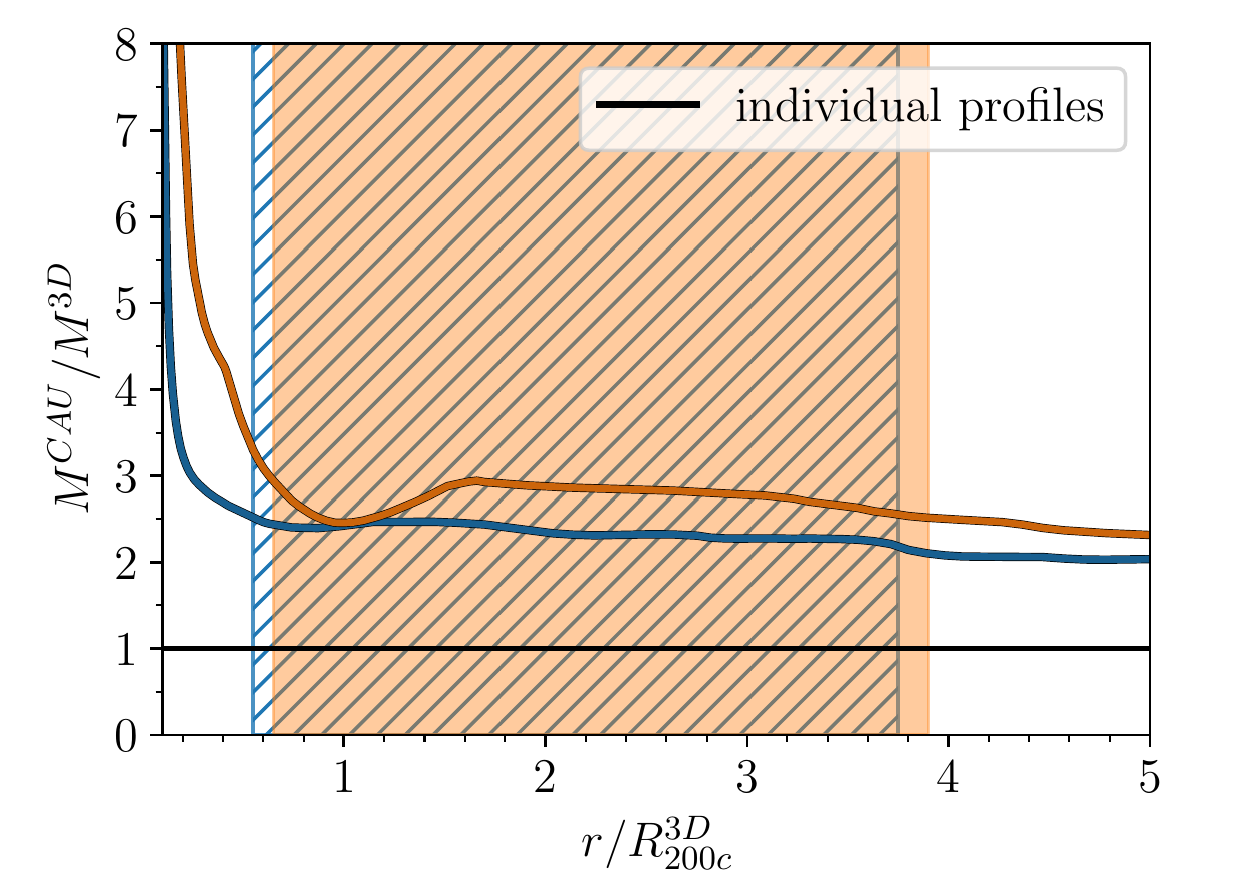}
    \caption{Ratio between the caustic and the 3D true mass profile as a function of the cluster-centric distance $r/R_{200c}^{3D}$, for the two clusters at $z=0.11$ considered in Fig. \ref{fig:comp90gal}. The blue and orange vertical bands delimit the plateaus of the  corresponding colored profiles. (See Sect. \ref{subsec:fbeta}). }
    \label{fig:plateau}
\end{figure}

In Fig. \ref{fig:plateau}, the two curves show the ratio between the caustic mass profile, $M^C$ (= $G^{-1} \int_{0}^{r} {\cal A}^2(R) \,{\rm d}R$),
and the true mass profile, $M^{3D}$, as a function of $r/R_{200c}^{3D}$ for two clusters at redshift $z=0.11$. 
The blue (orange) profile refers to the same cluster used for the top (bottom) $r-v_{\rm los}$ diagram in Fig. \ref{fig:comp90gal}.
The ratios show  three clear regimes. At small radii, $r\lesssim 1R_{200c}^{3D}$, there is a rapid decrease. At intermediate cluster-centric distances, $1R_{200c}^{3D} \lesssim r \lesssim 4R_{200c}^{3D}$, there is a  plateau. At larger radii, the ratios decrease. In Fig. \ref{fig:plateau}, the colored vertical bands delimit the plateau for the correspondingly colored individual profile.

On the plateau, with a typical extent of $\sim 3R_{200c}^{3D}$, the ratio between the caustic and the true mass profiles is approximately constant. Fig. \ref{fig:plateau} shows that the caustic mass profiles 
overestimate the true profiles in this radial range by a nearly constant factor. The mass ratio on the plateaus is a direct measure of the filling factor, $\mathcal{F}_\beta$, for each profile. 

The first step in measuring the filling factor is  the definition of the plateau. For each redshift sample, we compute the $M^{C}/M^{3D}(r/R_{200}^{3D})$ profiles for each cluster. We then bin the clusters in $N=101$ equally spaced bins of $r/R_{200c}^{3D}$ in the range $(0.1-5)R_{200c}^{3D}$. 

In building the distribution of the mass ratio profiles, we reject a cluster if its caustic amplitude shrinks to zero at a radius $ < 5R_{200c}^{3D}$ because the S/N ratio decreases substantially at large cluster-centric distances. On average we remove only 1.5\% of the sample for this reason.

For each redshift, we adopt an iterative procedure to identify the radial range of the plateau.
We begin by computing $\sigma_{0}$, the standard deviation of the mass ratios $M^{C}/M^{3D}$ over the entire radial range $(0.1-5)R_{200c}^{3D}$. At each iteration, $n$, we then compute two new standard deviations. We first omit the bin at the smallest radius from the sample and compute $\sigma_{n,b}$ for the remaining interval. Then we omit only the bin at the largest radius and compute $\sigma_{n,e}$ for all of the remaining bins. The subscripts $b$ and $e$ indicate omission of a bin at the ``beginning'' and at the ``end'' of the radial range, respectively. If $\sigma_{n,b} < \sigma_{n,e}$, we define $\sigma_n\equiv\sigma_{n,b}$ and drop the first radial bin. If $\sigma_{n,b} > \sigma_{n,e}$, $\sigma_n\equiv\sigma_{n,e}$ and we drop the last radial bin. Iterating this procedure for $n\leq N-1$ where N = 101 yields a set of shrinking radial intervals with the goal of minimizing  the dispersion in the distribution of mass ratios at each step.

As the interval is reduced, the standard deviations first decrease rapidly and then decreases slowly. The onset of the shallower decrease sets the radial limits of the plateau.
To locate the plateau numerically, we calculate the set of $\sigma_n$ for $i=0,...,N-10$ (where $N = 101$). We select ten sequential standard deviations $\sigma_{n},...,\sigma_{n+9}$ for iterations $n$ = 0, 1, 2, etc. Then we  calculate  nine differences $\delta \sigma_{n,i} = |\sigma_{n+i} - \sigma_{n+i+1}|$ where $i$ indicates the $i$th difference and the bars indicate the modulus of the difference. The average of the nine values of $\delta \sigma_{n,i}$ is $\braket{\delta \sigma_n}$. For the next set, $n$ + 1, the average is $\braket{\delta \sigma_{n+1}}$. 
We define the plateau as the largest interval (or, equivalently, lowest $n$) where the difference $\braket{\delta \sigma_n}$ is less than  $\varepsilon$. 
We choose $\varepsilon$ = 0.001. Our choice is based on the average behavior of $\braket{\delta \sigma_n}$ as a function of $n$. For the largest candidate plateaus (intervals), or, equivalently,  lowest values of $n$, $\braket{\delta \sigma_n}\sim 0.1$. For smaller intervals and larger  $n$ , $\braket{\delta \sigma_n}$ steadily decreases. The value reaches $\braket{\delta \sigma_n} \gtrsim 10^{-4}$ as the width of the remaining interval becomes $\lesssim 1R_{200c}^{3D}$. The choice  $\varepsilon = 0.001$ ensures  sufficient flatness over the plateau. Figure \ref{fig:plateau} shows the plateau limits for two of the clusters in the full sample.

In each redshift sample, the sequence of standard deviations may not relax below $\varepsilon$ for some clusters. We exclude these clusters in locating the  typical plateau. Depending on redshift, we exclude  14.5\% to 27.8\% of the clusters. On average we exclude 22.6\% of the sample. 

For each redshift, we compute a single plateau  delimited by the median of the smallest radius  and the median of the largest radius of the individual cluster plateaus. Finally, we compute a global plateau from the medians of the 11 smallest and the largest radii of the 11 plateaus at fixed redshift. This unique average plateau covers the  radial range $\mathcal{P}=(0.60-4.2)R_{200c}^{3D}$.

The second step in the calibration procedure exploits the plateau $\mathcal{P}$ to determine the filling factor as a function of redshift.
For  each cluster  in a redshift bin, we compute the average value of its profile ratio $\braket{M^{C}/M^{3D}(r/R_{200}^{3D})}$ on the plateau $\mathcal{P}$. 
On the interval $\mathcal{P}$, $\braket{M^{C}/M^{3D}(r/R_{200}^{3D})}^{-1}$ measures the  filling factor for that individual $k$th cluster, $\mathcal{F}_{\beta k}$ . 
The optimal filling factor for the cluster sample in a given bin is the average of the estimates for the individual clusters.

We measure $\mathcal{F}_{\beta k}$ for the clusters in each redshift sample. This procedure yields a distribution of $\mathcal{F}_{\beta k}$'s. 
\begin{figure*}[htbp]
    \includegraphics[width=\textwidth]{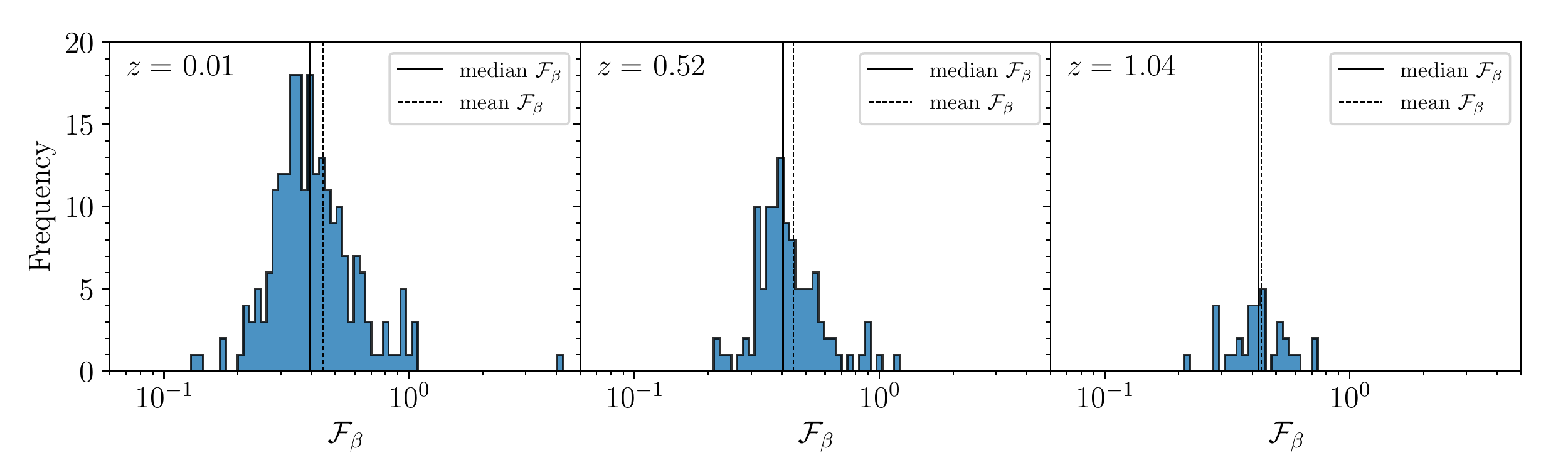}
    \caption{Histograms of the individual $\mathcal{F}_{{\beta} k}$ for 3  cluster samples at different redshifts. Redshift increases from left to right. The thick solid (thin dotted) line in each panel shows the median (mean) of the distribution.} \label{fig:fb_hist}
\end{figure*}
Figure \ref{fig:fb_hist} shows the distribution of individual $\mathcal{F}_{\beta k}$'s for samples at  three different redshifts ($z=0.01,0.52$, and $1.04$ from left to right, respectively). The distributions have an asymmetric bell shape that is skewed toward large values of $\mathcal{F}_{\beta}$. The mean of each distribution (dashed line) has a substantial offset with respect to the median (solid line). The means are sensitive to the tails of the distributions, but the medians are  located near the peak of the distributions. 
For each cluster sample, we take the median of the individual $\mathcal{F}_{\beta k}$'s as the optimal filling factor for the redshift interval.

\begin{figure}
    \centering
    \includegraphics[width=\columnwidth]{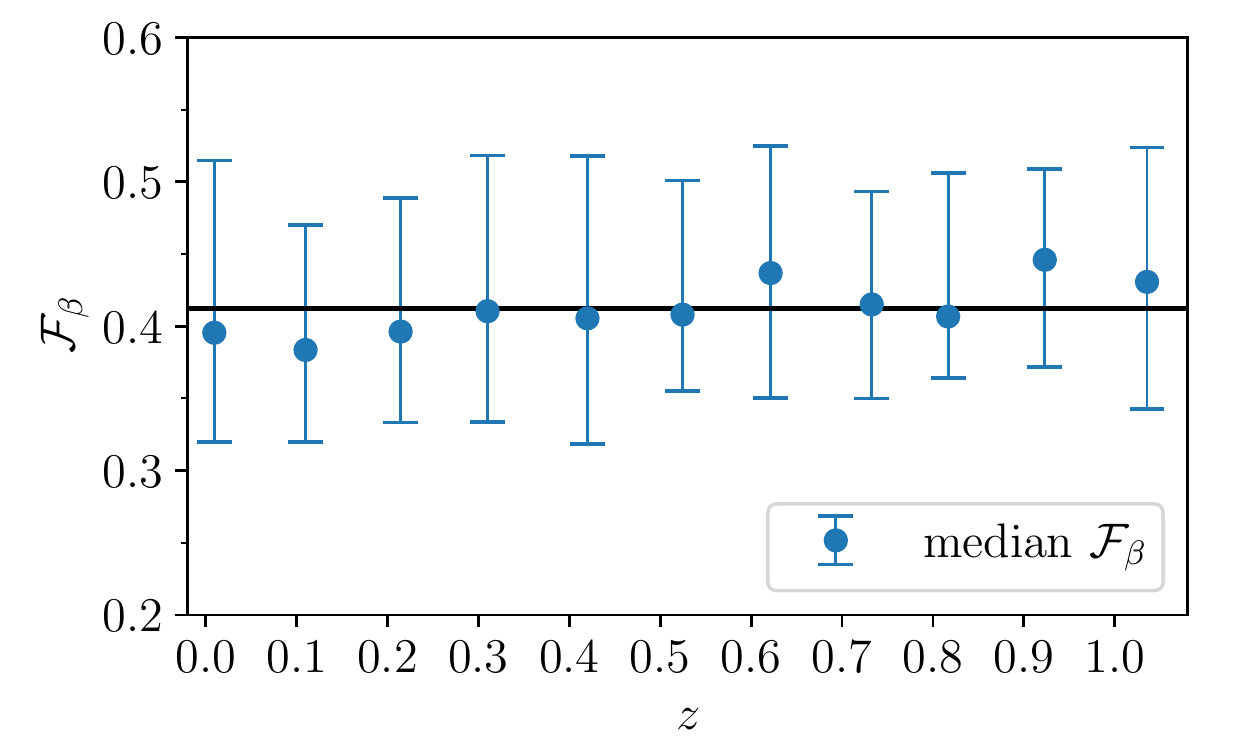}
    \caption{$\mathcal{F}_\beta$  as a function of redshift. Error bars show the interquartile range for each redshift sample. The black horizontal line shows the average value of the filling factor over redshift, $\bar{\mathcal{F}_\beta}=0.41$.}
    \label{fig:fb}
\end{figure}

Table \ref{table:fb}  lists  the numerical values of the filling factor and its interquartile range in the 11 redshift bins.
\begin{table}[htbp]
\begin{center}
\caption{\label{table:fb} Filling Factors} 
\begin{tabular}{cccc}
\hline
\hline
 $z$ & no of clusters & median $\mathcal{F}_\beta$ & $50\%$ range \\
     & &   & \\
\hline
 & & \\
0.01 & 224 & 0.40 & (0.32-0.51) \\
0.11 & 212 & 0.38 & (0.32-0.47) \\
0.21 & 181 & 0.40 & (0.33-0.49) \\
0.31 & 157 & 0.41 & (0.33-0.52) \\
0.42 & 134 & 0.41 & (0.32-0.52) \\
0.52 & 109 & 0.41 & (0.36-0.50) \\
0.62 & 97 & 0.44 & (0.35-0.52) \\
0.73 & 70 & 0.42 & (0.35-0.49) \\
0.82 & 57 & 0.41 & (0.36-0.51) \\
0.92 & 44 & 0.45 & (0.37-0.51) \\
1.04 & 33 & 0.43 & (0.34-0.52) \\
\hline
 \end{tabular}
 \end{center}
 \end{table}
Figure \ref{fig:fb} shows the optimal filling factor as a function of redshift. The error bars show the interquartile ranges. The horizontal line shows the mean filling factor  averaged over the 11 redshift ranges, $\bar{\mathcal{F}_\beta}=0.41$.
Figure \ref{fig:fb} and Table \ref{table:fb} show that the optimal filling factor is essentially constant throughout the entire redshift range that we consider. The factor is $\mathcal{F}_\beta\sim 0.40$ for $z<0.62$ and it increases by $\sim 7$\%  at higher redshifts. The increase with redshift is not statistically significant and is a result of the limited size of the cluster samples at $z\geq 0.62$. For $z> 0.62$ the samples contain fewer than 100 clusters (Table \ref{table:fb}, column 2). 
The $\tau$ statistic of Kendall's non-parametric measure of the correlation between $\mathcal{F}_\beta$ and redshift is $\sim 0.06$; thus we conclude that $\mathcal{F}_\beta$ is independent of redshift.
This redshift independence of $\mathcal{F}_\beta$ is a distinctive result of the calibration of the caustic technique with Illustris TNG300-1.  

\section{The caustic technique as a cluster mass profile estimator}\label{sec:performance}

Based on the optimal filling factor, we evaluate the  caustic technique as an estimator of the cluster characteristic radius and mass. For clusters where  $R_{200c}^{3D}$ lies within the interval $\mathcal{P}$, we compare the caustic  radius $R_{200c}^{C}$ with $R_{200c}^{3D}$. The ratio between the caustic and the true mass profile of the simulated clusters   provides the basis for use of the caustic technique as a method for determining cluster mass profiles in observational datasets. 

We begin by comparing the radii $R_{200c}^{C}$'s estimated from the caustic profiles with the true $R_{200c}^{3D}$'s. For each cluster in the simulation (Sect \ref{sec:fb}), we compute $R_{200c}^{C}$. We remove profiles where $R_{200c}^{C}$ is indeterminate because it lies at too small a radius to overlap the calibrated range $\mathcal{P}=(0.6-4.2)R_{200c}^{3D}$ where the technique holds. Fewer than 1\% of the systems have an indeterminate $R_{200c}^{C}$. 

\begin{figure}
    \centering
    \includegraphics[width=\columnwidth]{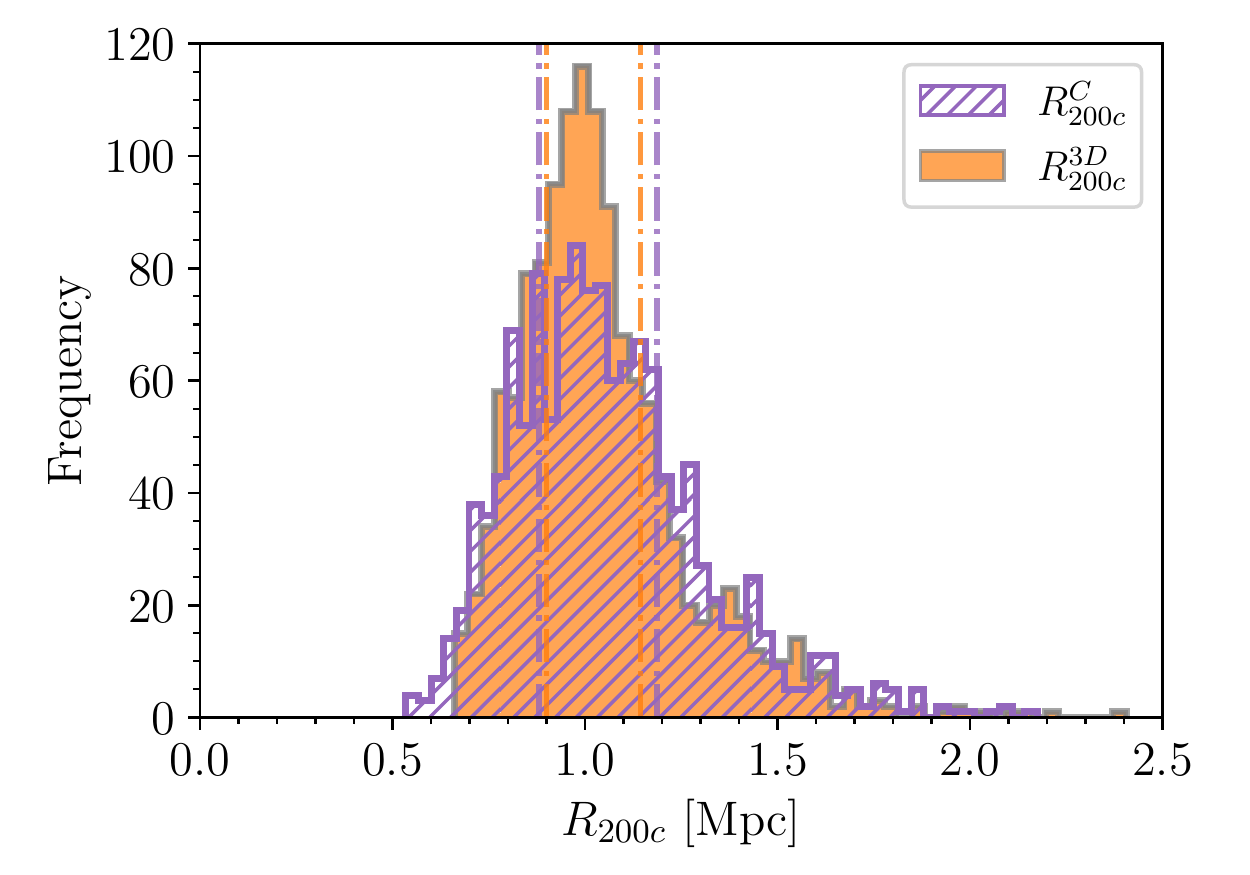}
    \caption{Histograms of $R_{200c}^{C}$ (violet) and $R_{200c}^{3D}$ (orange). The violet (orange) vertical lines show the interquartile ranges of the distributions of $R_{200c}^{C}$ ($R_{200c}^{3D}$).}
    \label{fig:r200}
\end{figure}
In Fig. \ref{fig:r200} the violet and orange histograms show the distributions of the caustic and true $R_{200c}$, respectively for all 11 redshift samples. The  dash-dotted vertical line shows the interquartile range of the distribution with  the corresponding color. The two distributions are similar.  The peaks of both distributions are at $\sim 1$~Mpc; the median $R_{200c}^{C}$ is only 1.9\% larger than the median $R_{200c}^{3D}$. Because $R_{200c}^{C}$ has a larger intrinsic error, the interquartile range of the caustic radii, $(0.88-1.19)$~Mpc slightly exceeds the interquartile range of the  true  3D radii, $(0.90-1.14)$~Mpc. 

\begin{figure}
    \centering
    \includegraphics[width=\columnwidth]{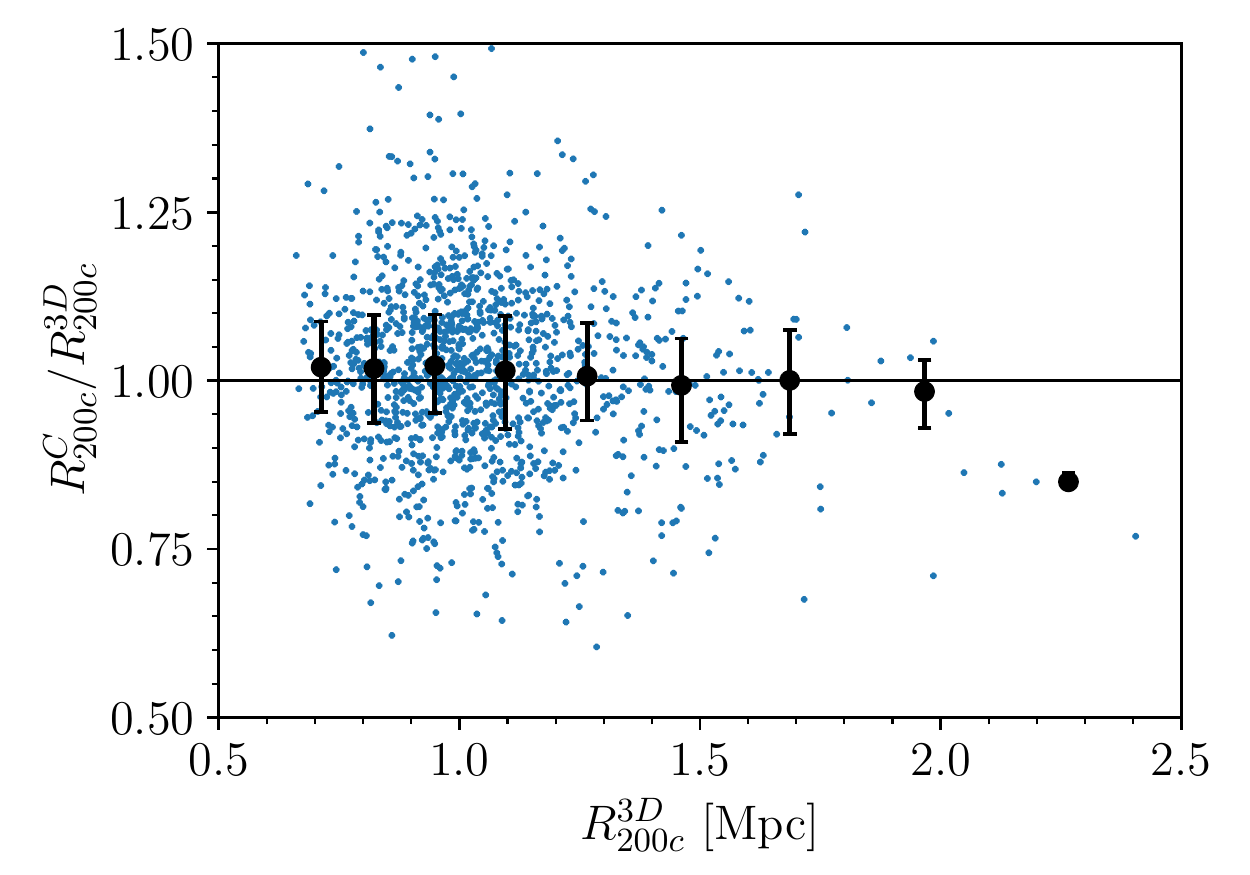}
    \caption{Ratios between $R_{200c}^C$ and $R_{200c}^{3D}$ as a function of $R_{200c}^{3D}$ for the  simulated clusters. We omit 12 clusters with a ratio $> 1.5$ for clarity. Points with error bars show the median and interquartile range of all the ratios in nine logarithmic bins of $R_{200c}^{3D}$. The horizontal line shows $R_{200c}^C=R_{200c}^{3D}$.}
    \label{fig:r200c3d}
\end{figure}
Figure \ref{fig:r200c3d} shows the ratio between the caustic and true $R_{200c}$'s as a function of $R_{200c}^{3D}$. Each point represents a simulated cluster in one of the 11 redshift samples. In the plot, we omit 12 clusters with a ratio $> 1.5$ to show the dispersion around a ratio of 1 more clearly.
The nine logarithmic bins spanning the range of $R_{200c}^{3D}$'s account for the relative sampling densities at the lower and upper extrema of true radii. 

In each bin, we compute the median and the interquartile range of all the ratios (Fig. \ref{fig:r200c3d},  black points with error bars). 
The horizontal line indicates $R_{200c}^C=R_{200c}^{3D}$. The ratios are generally consistent with equality, but the medians of the best sampled bins show that $R_{200c}^{C}$ overestimates $R_{200c}^{3D}$ slightly. The typical overestimation is $\sim 1.4\%$ and the dispersion is $\sim 6.6\%$. Thus the small overestimate is well within the error in its determination.

We  next assess the performance of the caustic technique as a mass estimator in the interval $\mathcal{P}$.
For each cluster in each sample, we normalise the caustic profile by $R_{200c}^{C}$ and obtain $M^C(r/R_{200c}^{C})$. Similarly, we normalise the true mass profile by the true $R_{200c}^{3D}$ to determine $M^{3D}(r/R_{200c}^{3D})$. We then compute the ratio of the two profiles as a function of  $r/R_{200c}^{C}=r/R_{200c}^{3D}$. Finally we compute the median profile of the distribution of individual mass ratio profiles. 

\begin{figure*}
    \centering
    \includegraphics[width=\textwidth]{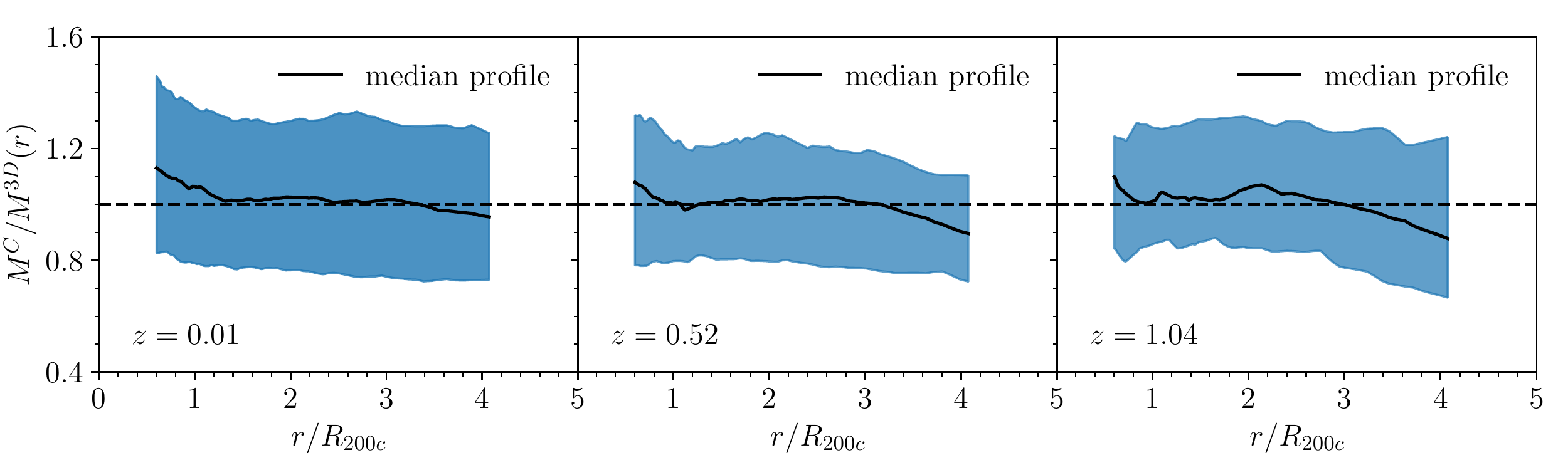}
    \caption{Ratio between the caustic and true mass profiles of the simulated clusters for  three different redshifts, $z=0.01, 0.52$, and $1.04$ (left to right). The solid black curve in each panel shows the median radial profile of the mass ratios. The blue area shows the interquartile range. The black dashed line shows  
    $M^{C}(r)=M^{3D}(r)$.}
    \label{fig:comp_mass_glob}
\end{figure*}
The solid black curves in the three panels of Fig. \ref{fig:comp_mass_glob} show the median mass ratio profile in the redshift ranges $z=0.01$, $z=0.52$, and $z=1.04$ (left to right). The shaded blue areas show the interquartile range of the profiles. The dashed horizontal line shows equality between the caustic and the true mass profile.

The median caustic and true profiles are on average equal in the  interval  $\mathcal{P}$ where we calibrate $\mathcal{F}_{\beta}$. Averaging over redshift, the median mass ratio in the calibrated range is unbiased and  lies between 0.92 and 1.10. On average the caustic technique returns unbiased mass profiles within 10\% of the true profiles. The average uncertainty is $\sim 23\%$, and is typically $\lesssim$ 29\%. At $R_{200c}$, on average $M^C/M^{3D}=1.03\pm 0.22$. At $2R_{200c}$, on average $M^C/M^{3D}=1.02\pm 0.23$.

\section{Discussion}\label{sec:discussion}
The Illustris TNG-300 simulations enable the first test of the caustic technique for cluster mass estimation based on  galaxies extracted from hydrodynamical simulations (Sect. \ref{sec:performance}). Previous analyses were based either on galaxies identified in N-body simulations with semi-analytical methods \citep{Diaferio99} or random samples of dark matter particles \citep{Serra2011}. In contrast with earlier work, the Illustris TNG300-1 results span a broad redshift range $0.01-1.04$. 

Illustris TNG-300 provides the basis for testing the caustic technique on both galaxies and dark matter for the same set of clusters. In Sect. \ref{subsec:caudm} we ask whether the simulated galaxies  in these clusters are biased tracers of the dark matter distribution. Sect. \ref{subsec:prev} surveys these results and compares them with Illustris TNG300-1. 

\subsection{Comparing the caustic technique for galaxies and dark matter}\label{subsec:caudm}

Galaxies, real or simulated, may be biased tracers of the dark matter distribution. To
examine this issue  we apply the caustic technique to the dark matter distribution in the same cluster catalogs we explore with simulated galaxies. 

For this test we use two catalogs: 255 simulated clusters at $z=0.11$ and  178 clusters at $z=0.42$ (Table \ref{table:3dinfo}). For each cluster, we build one dark matter mock redshift survey following the procedure outlined in Sect. \ref{subsec:mockgal}.  We use the Subfind dark matter substructures (rather than galaxies) with stellar mass larger than $10^7$~M$_\odot$. This stellar mass threshold encompasses more than 96\% of the total cluster substructures resolved by Subfind.
Among these substructures, we select those with total mass larger than $2.1\cdot 10^{10}$~M$_\odot$; the number of substructures within $3R_{200}^{3D}$ is then comparable to the number of galaxies in the corresponding galaxy catalogues. The difference between galaxy and dark matter catalogues is driven by the presence, in the latter, of dark matter subhalos with total mass comparable to galaxies.

The dark matter mock catalogues contain, on average,  $62\%$  more substructures than the corresponding  number of galaxies. The primary difference is that the dark matter mock catalogues are richer in foreground and background  structures than the corresponding galaxy catalogues.

We apply the caustic technique to the dark matter catalogues following Sect. \ref{subsec:caugal}. For each of the two sets of dark matter catalogues we choose $h_c$ as the median of the distribution of the individual $h_c$'s determined from the galaxy mock catalogues at the corresponding redshift (Sect. \ref{subsec:caugal}). In other words, we locate the caustics from the 2D projected phase-space dark matter densities with the smoothing scale adopted for galaxies. The smoothing parameters are $h_c=0.44$ and $h_c=0.50$ for $z=0.11$ and $z=0.42$, respectively. 

We calibrate the filling factor from the dark matter caustic profiles following the  procedure  in Sect. \ref{subsec:fbeta}.
First, we locate the common plateau of the mass profiles from the dark matter catalogues. In this procedure, we end up  removing 19.3\% of cluster profiles where the plateau is indeterminate. This percentage is analogous to the 22.6\% removal of systems based on simulated galaxy catalogs. 

The dark matter caustic profiles yield a well defined global plateau with a radial range $(0.45-4.7)R_{200c}^{3D}$. This plateau extends to larger cluster-centric distances than the plateau based on the analogous galaxy catalogues. The more extended dark matter plateau results from the excess of dark matter substructures compared with simulated galaxies. For the galaxy catalogs the extent of the plateau is $\mathcal{P}=(0.60-4.15)R_{200c}^{3D}$ and it is included within the dark matter plateau. We limit the analysis of dark matter caustic profiles to the plateau region defined by the simulated galaxies.

We measure the filling factor from the dark matter caustic profiles following Sect. \ref{subsec:fbeta}. At $z=0.11$, $\mathcal{F}_\beta=0.37$ with an interquartile range $(0.30-0.46)$. At $z=0.42$, $\mathcal{F}_\beta=0.37$ with  an interquartile range $(0.31-0.44)$. These values are in excellent agreement with those obtained from galaxy mock catalogues (Table \ref{table:fb}).

\begin{figure}
    \centering
    \includegraphics[width=\columnwidth]{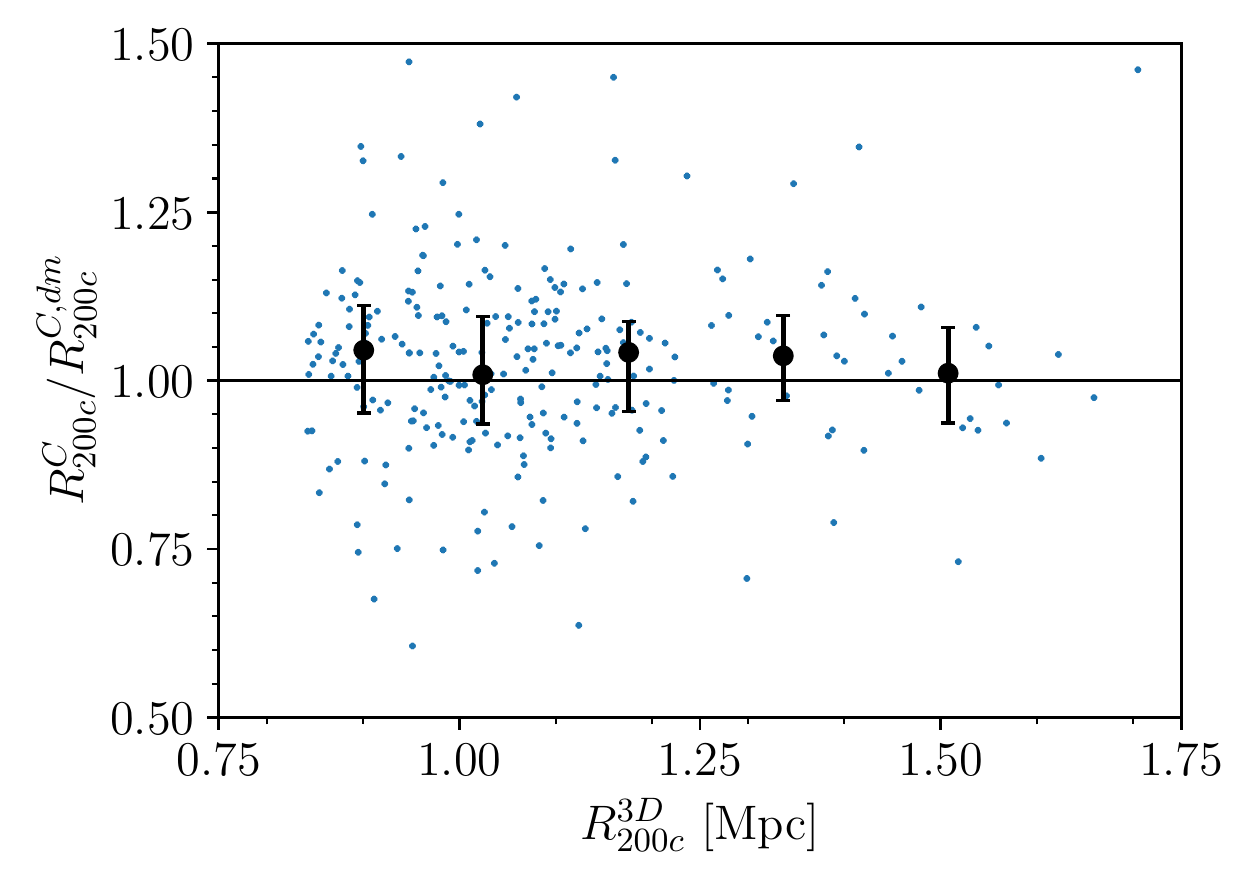}
    \caption{$R_{200c}^C /R_{200c}^{C,dm}$ as a function of $R_{200c}^{3D}$ for simulated clusters at $z=0.11$ and $z=0.42$ (blue points). Black points with error bars show the median and the interquartile range of all the ratios in five logarithmic bins of $R_{200c}^{3D}$. The horizontal line is $R_{200c}^C=R_{200c}^{C,dm}$. }
    \label{fig:r200_galdm}
\end{figure}

We compare the radii $R_{200c}^C$ estimated from the galaxy caustic profiles with the corresponding set of $R_{200c}^{C,dm}$ estimated from the dark matter caustic profiles. There are 272 clusters that allow estimation of both radii on the plateau $\mathcal{P}$. Fig. \ref{fig:r200_galdm} shows the ratio $R_{200c}^{C}/R_{200c}^{C,dm}$ for each cluster (blue points) in the two redshift samples. 
Black points with error bars show the median and interquartile range of the distribution of these ratios in five logarithmic bins of $R_{200c}^{3D}$. In the plot, we omit 4 clusters with ratios larger than 1.5 and the 4 clusters with $R_{200c}^{3D}$ larger than $1.75$~Mpc from the plot. These systems have a negligible effect on the median and the interquartile range in the relevant bins. On average, the radii based on simulated galaxies slightly exceed the radii based on dark matter by $\sim 3.3\%$. The typical interquartile range  is $\sim 7\%$. The small overestimation is thus well within the uncertainty.

The ratio  between the simulated galaxy and dark matter caustic mass profiles of individual clusters provides a platform for  assessing the bias between the dark matter and simulated galaxies as tracers  of the matter distribution derived from the caustic technique. There are 177 and 110 clusters that support this analysis at $z=0.11$ and $z=0.42$, respectively.

We normalise the galaxy and the dark matter based caustic mass profiles of each cluster by its $R_{200c}^{3D}$, and obtain $M^{C}(r/R_{200c}^{3D})$ and $M^{C,dm}(r/R_{200c}^{3D})$, respectively. Then, for each cluster, we compute the ratio profile of the normalised galaxy-based and dark-matter-based mass profiles at fixed $r/R_{200c}^{3D}$. 

\begin{figure}
    \centering
    \includegraphics[width=\columnwidth]{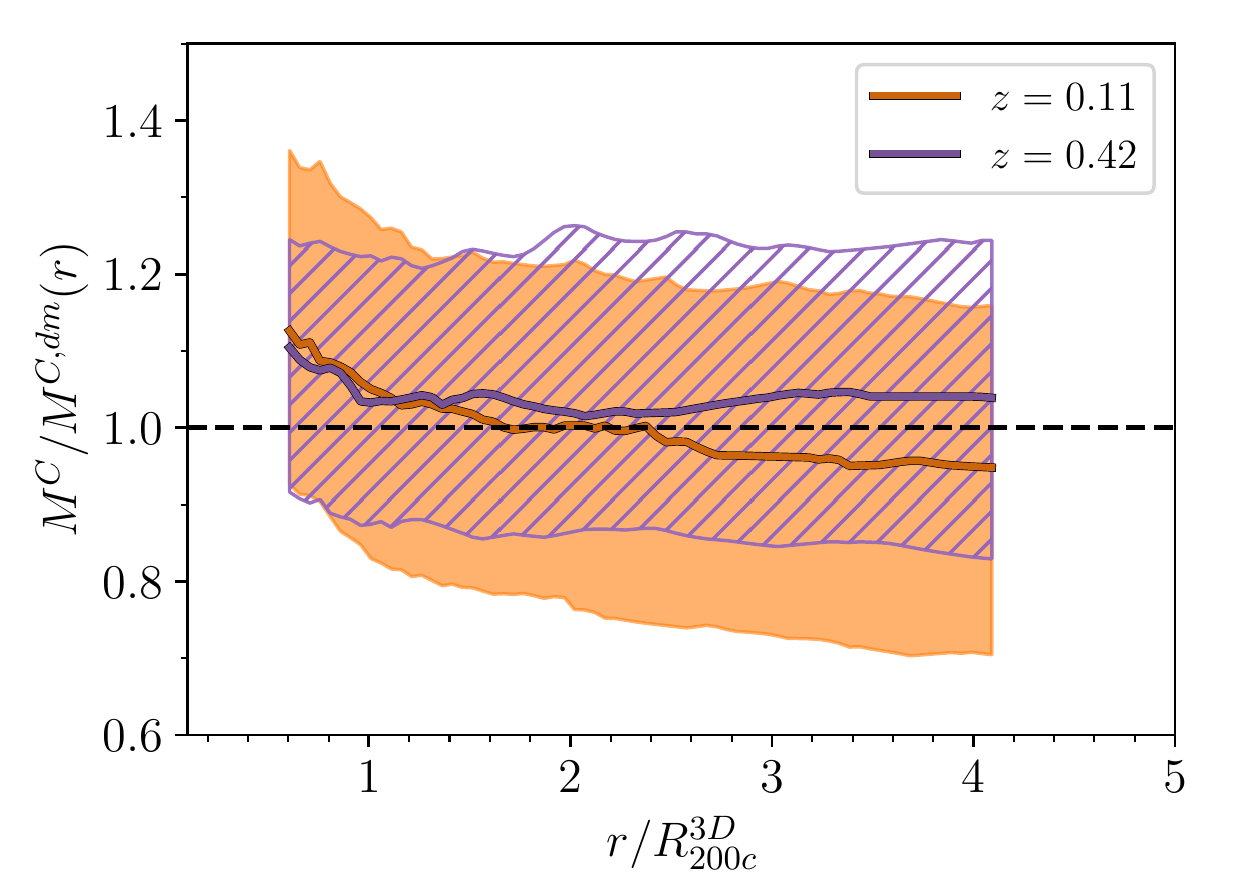}
    \caption{Ratio between caustic profiles based on simulated galaxies and dark matter, for  clusters at $z=0.11$ (orange) and $z=0.42$ (violet). Solid curves show the median radial profile of the mass ratios. The shaded regions areas show the interquartile range. The black dashed line shows $M^{C}(r)=M^{C,dm}(r)$. }
    \label{fig:fb_dml}
\end{figure}

Fig. \ref{fig:fb_dml} shows the median of the individual ratio profiles for $z=0.11$ (orange) and $z=0.42$ (violet). The shaded areas show the interquartile ranges of the profiles. At $z=0.11$, $M^{C}$ is $\sim 2\%$ smaller than $M^{C,dm}$; at $z=0.42$, $M^{C}$ is $\sim 4\%$ larger than $M^{C,dm}$. The difference in the ratio is small compared with the interquartile range of $\sim$ 21\%. Thus comparison of the caustic technique applied to both galaxy and dark matter catalogs in Illustris TNG 300-1 demonstrates that the simulated galaxies are essentially unbiased tracers of the dark matter distribution.  

\subsection{Comparison with previous investigations}\label{subsec:prev}

We next place the IllustrisTNG results in the context of previous investigations of the caustic technique.  In particular, we review various estimates of the filling factor $\mathcal{F}_\beta=0.5$. We briefly review observational applications of the technique and preview future  extensions of the caustic technique to large cluster redshift survey that extend to high redshift. Throughout we emphasize the robustness and redshift independence of $\mathcal{F}_\beta=0.41$ demonstrated by IllustrisTNG.

The caustic technique  was originally developed by \citet{Diaferio1997} and \citet {Diaferio99}. Other studies of the technique \citep[e.g.][]{biviano2003mass,Serra2011,Gifford13a,Gifford13b} adopt a variety of complementary technical approaches including  variations in the algorithm for locating the caustics. They may also adopt non-constant filling factors $\mathcal{F}_\beta(r)$. We limit our detailed comparisons to the work of \citet{Diaferio1997} and \citet{Diaferio99}, where the approach is most similar. Most observational analyses employ the the caustic technique based on this work. 

Initially \citet{Diaferio1997} derived $\mathcal{F}_\beta=0.5$ assuming a hierarchical clustering scenario. \citet{Diaferio99} used the  GIF \citep{Kauffmann99} $\Lambda$CDM N-body simulation at $z=0$ to provide the first simulation-based evaluation of $\mathcal{F}_\beta(r)$. This estimate is based on the cluster mass density, the potential, and the velocity anisotropy (see Sect. \ref{sec:ct}). The value $\mathcal{F}_\beta=0.5$ is an average value of $\mathcal{F}_\beta(r)$ within $\sim (1-3)R_{200c}$.

The IllustrisTNG calibration, $\mathcal{F}_\beta=0.41$, is typically $\sim 18$\% smaller than previous results, but it is based on a broader, more robust platform. The earlier studies are based on collisionless N-body simulations with  more limited volume and with lower resolution. In contrast TNG300-1 \citep{Pillepich18,Springel18,Nelson19} is a benchmark for large-scale hydrodynamical simulations. 
The Illustris TNG300-1 measurement of the filling factor is based only on the relationship between the caustic and the true mass profile of clusters evaluated consistently from the simulated clusters. In contrast with previous work, we extend the measurement of the filling factor beyond $z \sim 0$ and reach a limiting $z\sim 1$. 

Illustris TNG300-1 is a platform for establishing a standardised statistical approach for application of the caustic technique to the  outskirts of clusters of galaxies.  
The identical, unconstrained setup of the analysis applies to every optimally sampled cluster (Sect. \ref{subsec:caugal}). Larger volume hydrodynamical simulations like MillenniumTNG will enable the calibration of the caustic technique over the full observed cluster mass range. These larger simulations will naturally include a larger number of the most massive systems than Illustris TNG300-1. These simulations will enable evaluation of the sensitivity of the caustic method to the cluster mass, an analysis that is not feasible with Illustris TNG300-1.

Comparisons between caustic and weak lensing masses \citep{diaferio2005caustic,Geller_2013} and caustic and X-ray masses \citep[e.g.,][]{Maughan2016,Andreon2017,Lovisari20a,Logan22} generally show consistency between the caustic masses obtained with the implementation of \citet{Diaferio1997} and \citet{Diaferio99} for $\mathcal{F}_\beta=0.5$. Both the caustic technique and weak lensing have the strength that they are independent of equilibrium assumptions in contrast with X-ray estimates.

Weak lensing and the caustic technique probe a similar radial range of the cluster mass profile. In contrast,  X-ray approaches apply to the  smaller range $\sim (0-1)R_{200c}$.  In general, the spectroscopic sampling and the variable parameters used to analyze each cluster  make it difficult to assess the detailed reasons for differing results.

\citet{Geller_2013} show that the caustic mass estimate exceeds the weak lensing estimate by $\sim 20-30\%$ at radii $\sim (0.5-1.3)R_{200c}$ and is  $\sim 20-30\%$ {\it below} the weak lensing estimate in the radial range $\sim (1.3-3)R_{200c}$. On average  X-ray $M_{200c}$'s generally exceed the caustic masses by $\sim 10-30\%$ \citep{Maughan2016,Lovisari20a,Logan22}. \citet{Andreon2017} infers caustic masses $\sim$ 10\% larger than X-ray masses.

The $\sim 18\%$ smaller  (a caustic amplitude correspondingly lower by $\sim$9\%)  Illustris TNG300-1 filling factors reduce the caustic mass estimates by approximately this factor.  Taken at face value,  this revision of the filling factor implies underestimation relative to weak lensing  mass profile at large radii by $\sim 35-40\%$. Relative to the X-ray $M_{200c}$, the revised caustic masses are lower by $\sim 25-45\%$. These simple estimates ignore underlying differences between the Illustris TNG300-1 platform and the application of the caustic technique to previously observed cluster samples.

Optimal sampling of the cluster velocity field is  fundamental to robust application of the caustic technique. Redshifts of $\gtrsim 200$  members within a 3D radius of $3R_{200c}^{3D}$ \citep[e.g.][]{Serra2011} are optimal. Sparser sampling leads to smaller caustic amplitudes and thus an underestimate of the mass profile.

The  CIRS and HeCS survey \citep{Rines2006CIRS,Rines2013HeCS,sohn2019velocity} samples provide the best  presently available spectroscopic samples. There are typically $\lesssim 100-150$ total spectroscopic members within the probable $3R_{200c}^{3D}$.

Published observational studies already reflect the dependence  of the caustic masses on spectroscopic sampling. For example, out of the 19 clusters in the weak lensing comparison by \citet{Geller_2013}, the caustic masses of the 4 best sampled clusters exceed the  weak lensing mass by 20\% and up to 50\% over the entire radial range. This behavior contrasts with the average underestimate of the caustic masses relative to weak lensing masses.

\citet{Lovisari20a} compute the ratio between the X-ray mass $M_{500c}^{X}$ and the caustic $M_{500c}^{C}$ of 25 clusters. On average the ratio $M_{500c}^{X}/M_{500c}^{C}$ decreases from 1.30 to 1.03 as the sampling increases. \citet{Logan22} also show that $M_{500c}^{X}/M_{500c}^{C}$ decreases (from $\sim 1.69$ to $\sim 1.12$) as the average number of cluster members increases from 93 to 181. This effect results from the natural underestimate of the caustic amplitude as the sampling becomes poorer.

The Illustris TNG300-1 platform for the caustic technique  has no fine tuning of the parameters in the global analysis.  Previous application of the caustic technique includes fine tuning of various parameters fundamental to caustic mass estimates. These parameters modify the sample of candidate caustic members, the smoothing parameter, and/or  the selection of the isocurve in the continuous projected phase-space of the cluster galaxies. These effects can easily exceed the $\sim 9\%$ reduction of the caustic amplitude in the Illustris TNG-300 calibration relative to previous results.

The Illustris TNG-300 investigation of the caustic technique provides robust estimates of cluster mass profiles  covering  a significant radial range  extending beyond the virialised region of clusters. The calibration of the caustic technique is stable over a large redshift range, $z=0.01-1.04$.
 Upcoming large samples of  spectroscopically observed clusters will provide a basis for application of the Illustris TNG300-1  approach to caustic mass estimation. Planned observations with   the William Herschel Telescope Enhanced Area Velocity Explorer on WHT \citep[WEAVE,][]{Dalton12}, and the Prime Focus Spectrograph on Subaru \citep[PSF,][]{Tamura16}  and eventually the Maunakea Spectroscopic Explorer on CFHT \citep[MSE,][]{MSE19}, should provide thousands of clusters at redshifts $\sim 0.5-0.6$ including spectroscopically determined  redshifts of hundreds of members.  

\section{Conclusion}\label{sec:conclusion}

The IllustrisTNG simulations open a new window on the application of the caustic technique for measuring the masses of clusters of galaxies \citep{Diaferio1997,Diaferio99}. From the TNG300-1 simulation, we construct a catalogue of 1697 optimally sampled clusters with  $ M > 10^{14}$M$_\odot$ \citep{Pillepich18,Springel18,Nelson19} that cover the redshift range $0.01-1.04$. We then derive the total mass including dark matter, stars, gas, and black holes as a function of cluster-centric distance $r$ for each cluster. After applying the caustic technique to each cluster, we calculate the ratio of caustic mass to the total 3D mass, $\mathcal{F}_\beta$, as a function of $r$ and look for the span of $r$ where $\mathcal{F}_\beta$ is roughly constant. 

The analysis yields a clear plateau where $\mathcal{F}_\beta$ = $0.41 \pm 0.08$ over a large range in cluster-centric distance, (0.6--4.2)~$R_{200c}$. The existence of this plateau enables robust calibration and application of the caustic technique to real systems of galaxies. The technique also provides an unbiased estimate of the true $R_{200c}^{3D}$, $R_{200c}^C/R_{200c}^{3D} = 1.014 \pm 0.066$. The range of the plateau, the derived $\mathcal{F}_\beta$, and the estimate for $R_{200c}$ are insensitive to redshift. The caustic technique returns unbiased mass profiles with an average uncertainty of 23\%. At $R_{200c}$, on average $M^C/M^{3D}=1.03\pm 0.22$; at $2R_{200c}$, on average $M^C/M^{3D}=1.02\pm 0.23$.

This approach has several distinct advantages over previous investigations. The simulations provide the phase space distributions of cluster galaxies and dark matter for the same systems. The analysis applies a single, unconstrained setup to every cluster to derive the caustic mass profile. We also 
employ a robust algorithm to find the plateau where $\mathcal{F}_\beta$ is roughly constant. The applicable redshift range for this technique, $z \lesssim 1$, is much larger than previous efforts that focus on $z \approx$ 0.

The caustic amplitude as a function of cluster-centric radius provides an estimate of the local escape velocity. For both galaxies and dark matter in the TNG300-1 simulations, the analysis yields a robust comparison between the mass profiles of both galaxies and dark matter simultaneously. This comparison demonstrates that galaxies are unbiased tracers of the mass distribution.

IllustrisTNG provides a broad statistical platform for application of the caustic technique to large samples of clusters with spectroscopic redshifts for $\gtrsim 200$ members. Instruments including WEAVE on WHT \citep{Dalton12}, PSF on Subaru \citep{Tamura16}, and eventually  MSE on CFHT \citep{MSE19} will provide the necessary observational foundation. These observations will support extensive comparisons between caustic masses and weak lensing masses extending to large radius. These two approaches to cluster mass estimation share the strength that they are independent of equilibrium assumptions and they extend outside the virial region. The large future datasets will also support application of the caustic technique as a route for measuring the cluster growth rate as a function of redshift \citep{diaferio2004outskirts,adhikari2014,More2015,deBoni2016,Diemer2017sparta2,walker2019physics,cataneo2020tests,pizzardo2020,Pizzardo2022}. The cluster mass accretion rate complements other techniques for measuring the growth rate of structure in the universe. 

\begin{acknowledgements}
We thank Jubee Sohn for useful discussions. 
M.P. and I.D. acknowledge the support of the Canada Research Chair Program and the Natural Sciences and Engineering Research Council of Canada (NSERC, funding reference number RGPIN-2018-05425).
The Smithsonian Institution supports the research of M.J.G. and S.J.K. A.D. acknowledges partial support from the INFN grant InDark.
Part of the present analyses was performed with the computer resources of INFN in Torino and of the University of Torino.
This research has made use of NASA's Astrophysics Data System Bibliographic Services.\\
All of the primary TNG simulations have been run on the Cray XC40 Hazel Hen supercomputer at the High Performance Computing Center Stuttgart (HLRS) in Germany. They have been made possible by the Gauss Centre for Supercomputing (GCS) large-scale project proposals GCS-ILLU and GCS-DWAR. GCS is the alliance of the three national supercomputing centres HLRS (Universitaet Stuttgart), JSC (Forschungszentrum Julich), and LRZ (Bayerische Akademie der Wissenschaften), funded by the German Federal Ministry of Education and Research (BMBF) and the German State Ministries for Research of Baden-Wuerttemberg (MWK), Bayern (StMWFK) and Nordrhein-Westfalen (MIWF). Further simulations were run on the Hydra and Draco supercomputers at the Max Planck Computing and Data Facility (MPCDF, formerly known as RZG) in Garching near Munich, in addition to the Magny system at HITS in Heidelberg. Additional computations were carried out on the Odyssey2 system supported by the FAS Division of Science, Research Computing Group at Harvard University, and the Stampede supercomputer at the Texas Advanced Computing Center through the XSEDE project AST140063.
\end{acknowledgements}

\bibliographystyle{aa}
\bibliography{michele}

\end{document}